\documentclass[prl,twocolumn,groupedaddress,amsmath,amssymb,showpacs,superscriptaddress]{revtex4}
\usepackage{graphicx}
\usepackage{xspace}
\usepackage{gensymb}
\usepackage{units}

\def\Ca{\rm{^{40}Ca^{+}}}
\newcommand{\ket}[1]{|#1\rangle}

\newcommand{\abs}[1]{|#1|}

\begin{document}


\title{Absolute frequency measurement of the $^{40}$Ca$^+$ $4s\ ^2S_{1/2}-3d\ ^2D_{5/2}$ clock transition}

\author{M. Chwalla}
\affiliation{Institut f\"ur Experimentalphysik, Universit\"at Innsbruck, Technikerstrasse 25, A--6020 Innsbruck, Austria}
\author{J. Benhelm}
\affiliation{Institut f\"ur Experimentalphysik, Universit\"at
Innsbruck, Technikerstrasse 25, A--6020 Innsbruck, Austria}
\affiliation{Institut f\"ur Quantenoptik und Quanteninformation
der \"Osterreichischen Akademie der Wissenschaften, Technikerstrasse
21a, A--6020 Innsbruck, Austria}
\author{K. Kim}
\thanks{Present adress: Department of Physics, University of Maryland, College Park, Maryland, 20742, USA}
\affiliation{Institut f\"ur Experimentalphysik, Universit\"at
Innsbruck, Technikerstrasse 25, A--6020 Innsbruck, Austria}
\author{G. Kirchmair}
\affiliation{Institut f\"ur Experimentalphysik, Universit\"at
Innsbruck, Technikerstrasse 25, A--6020 Innsbruck, Austria}
\affiliation{Institut f\"ur Quantenoptik und Quanteninformation
der \"Osterreichischen Akademie der Wissenschaften, Technikerstrasse
21a, A--6020 Innsbruck, Austria}
\author{T. Monz}
\affiliation{Institut f\"ur Experimentalphysik, Universit\"at Innsbruck, Technikerstrasse 25, A--6020 Innsbruck, Austria}
\author{M. Riebe}
\affiliation{Institut f\"ur Experimentalphysik, Universit\"at Innsbruck, Technikerstrasse 25, A--6020 Innsbruck, Austria}
\author{P. Schindler}
\affiliation{Institut f\"ur Experimentalphysik, Universit\"at Innsbruck, Technikerstrasse 25, A--6020 Innsbruck, Austria}
\author{A. S. Villar}
\affiliation{Institut f\"ur Experimentalphysik, Universit\"at Innsbruck, Technikerstrasse 25, A--6020 Innsbruck, Austria}
\author{W. H\"ansel}
\affiliation{Institut f\"ur Experimentalphysik, Universit\"at Innsbruck, Technikerstrasse 25, A--6020 Innsbruck, Austria}
\author{C. F. Roos}
\email[]{christian.roos@uibk.ac.at}
\affiliation{Institut f\"ur Experimentalphysik, Universit\"at
Innsbruck, Technikerstrasse 25, A--6020 Innsbruck, Austria}
\affiliation{Institut f\"ur Quantenoptik und Quanteninformation
der \"Osterreichischen Akademie der Wissenschaften, Technikerstrasse
21a, A--6020 Innsbruck, Austria}
\author{R. Blatt}
\affiliation{Institut f\"ur Experimentalphysik, Universit\"at Innsbruck, Technikerstrasse 25, A--6020 Innsbruck, Austria}
\affiliation{Institut f\"ur Quantenoptik und Quanteninformation der \"Osterreichischen Akademie der Wissenschaften, Technikerstrasse 21a, A--6020 Innsbruck, Austria}
\author{M. Abgrall}
\affiliation{LNE-SYRTE, Observatoire de Paris, 61, Avenue de
l'Observatoire, 75014 Paris, France}
\author{G. Santarelli}
\affiliation{LNE-SYRTE, Observatoire de Paris, 61, Avenue de
l'Observatoire, 75014 Paris, France}
\author{G. D. Rovera}
\affiliation{LNE-SYRTE, Observatoire de Paris, 61, Avenue de
l'Observatoire, 75014 Paris, France}
\author{Ph. Laurent}
\affiliation{LNE-SYRTE, Observatoire de Paris, 61, Avenue de
l'Observatoire, 75014 Paris, France}

\date{\today}

\begin{abstract}

We report on the first absolute transition frequency measurement at the
$10^{-15}$ level with a single, laser-cooled $^{40}$Ca$^+$ ion in
a linear Paul trap. For this measurement, a frequency comb is
referenced to the transportable Cs atomic fountain clock of
LNE-SYRTE and is used to measure the $^{40}$Ca$^+$ $4s\
^2S_{1/2}-3d\ ^2D_{5/2}$ electric-quadrupole transition frequency.
After the correction of systematic shifts, the clock transition
frequency $\nu_{Ca^+} = 411\,042\,129\,776\,393.2 (1.0)$~Hz is
obtained, which corresponds to a fractional uncertainty within a
factor of three of the Cs standard. Future improvements are
expected to lead to an uncertainty surpassing the best Cs
fountain clocks. In addition, we determine the Land\'e g-factor of
the $3d\ ^2D_{5/2}$ level to be g$_{5/2}$=1.2003340(3).
\end{abstract}




\maketitle

Fostered by the advent of the optical frequency comb technology,
there is currently a strong trend towards the development of
optical frequency standards that are expected to replace the
current microwave standard in cesium as the basis of the
definition of the SI second. Recently, the best optical frequency
standards, based on single trapped ions and neutral atoms held in
optical lattices respectively \cite{rosenbandneu,ludlow}, have
both demonstrated a fractional frequency uncertainty of $10^{-16}$
or even better, thus surpassing the best cesium fountain clocks.
The great attraction of optical frequency standards lies in the
superior resonance line quality factors, allowing shorter
averaging times and higher stability. Experiments with single,
trapped ions have provided key contributions to the field of
optical clocks and precision measurements in the past 20 years
\cite{Madej01}. Several candidates for an optical ion clock have
been investigated such as Hg$^+$, Al$^+$, Yb$^+$, In$^+$, Sr$^+$
\cite{oskay,rosenband,stenger,blythe,indium,margolis}, or proposed
like Ca$^+$ \cite{champenois,kajita}.

\begin{figure}
\includegraphics[scale=0.95]{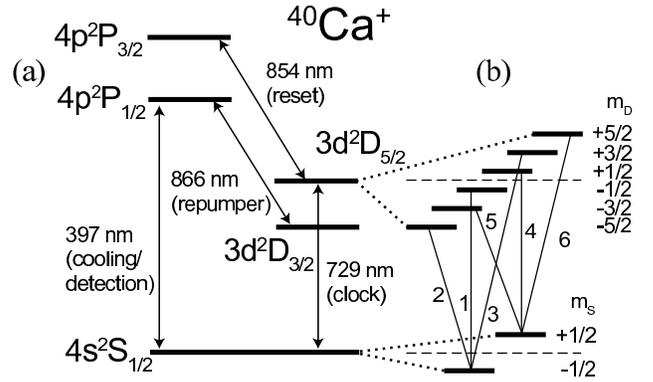}
\caption{\label{term} (a) Partial level scheme of $^{40}$Ca$^+$
showing the $S_{1/2}-D_{5/2}$ clock transition as well as the
transitions used for resetting, repumping, cooling, and detecting
the ion. (b) Magnetic sub-levels of the clock transition and the
six transitions having the biggest coupling strength for the
actual laser polarization, quantization axis, and k-vector of the
laser at 729 nm. The numbers indicate the order in which the
transitions were probed.}
\end{figure}

Building an ion clock based on Ca$^+$ has the technological
advantage that all necessary wavelengths for laser cooling and
state manipulation including lasers for photo-ionization can be
generated by commercially available and easy-to-handle solid state
lasers. Fig.\ \ref{term} shows the relevant atomic levels of the
isotope $^{40}$Ca$^+$. The actual clock transition considered here
is the electric quadrupole transition from the $4s\ ^2S_{1/2}$ to
the $3d\ ^2D_{5/2}$ level, which has a natural lifetime of
\unit[1.17]{s} \cite{barton}. A non-zero magnetic field lifts the
Zeeman degeneracy and splits the transition into ten lines.
Probing several of them provides a means to cancel the most
important systematic effects like the linear Zeeman and the
electric quadrupole shift \cite{margolis}.

\begin{figure}
\includegraphics[scale=0.65]{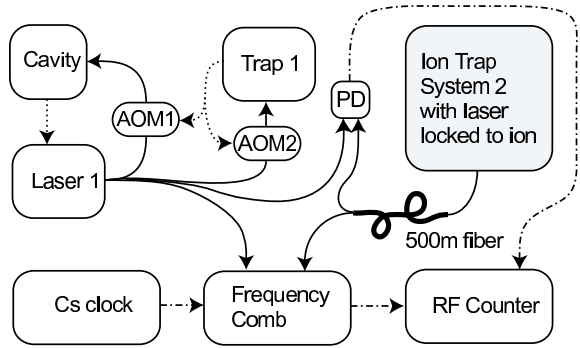}
\caption{\label{setup} Experiment setup: The frequency of an
ultra-stable laser is locked to a single $\Ca$ stored in a linear
ion trap and measured by a frequency comb referenced to a
transportable Cs fountain atomic clock. The light of a second
laser locked in a similar manner to another trap system in another
lab is transported via a \unit[500]{m} long optical fiber link. An
optical beat note is detected on a fast photodiode (PD). The solid
lines are optical signals, the dotted lines indicate electronic
feedback and the dashed-dotted lines show RF signals.}
\end{figure}

An overview of our experimental setup is given in
Fig.~\ref{setup}. We use a linear Paul trap \cite{setup} with four
blades separated by 2~mm and two tips of 5~mm separation providing
radial and axial confinement. Applying a radio-frequency (RF)
power of \unit[9]{W} to the trap and setting the tips to a
potential of \unit[1000]{V}, we achieve typical secular trap
frequencies $\omega_r/2\pi$=\unit[3.9]{MHz} in the radial and
$\omega_a/2\pi$=\unit[1.2]{MHz} in the axial direction. Single
Ca$^+$ ions are loaded into the trap by photo-ionizing a beam of
neutral calcium atoms.

A frequency measurement cycle consisted of \unit[2]{ms} of Doppler
cooling on the $S_{1/2} - P_{1/2}$ at a wavelength of
\unit[397]{nm} and repumping at \unit[866]{nm} and \unit[854]{nm}
to prevent pumping into the D$_{3/2}$ level and to clear out
population from the D$_{5/2}$ level. State initialization was
achieved by optical pumping into $\ket{S_{1/2},m_S=-1/2}$. After
that, laser pulses at \unit[729]{nm} from a titanium-sapphire
laser, which was locked to a high finesse cavity similar in design
to the one described in ref.~\cite{notcutt}, were used to probe
the $S_{1/2} - D_{5/2}$ transition. From an optical beat note with
a similar laser system we infer a typical linewidth (FWHM) of
\unit[50]{Hz} over 40 minutes. A constant magnetic field of
\unit[3.087(2)]{G} splits the Zeeman multiplet into ten
components, six of which ($\Delta m$=0 or $\abs{\Delta m}$=2)
could be excited with good coupling strength in the geometry
chosen for the laser polarization, k-vector, and magnetic field
\cite{hartmut}. To probe transitions starting from
$\ket{S_{1/2},m_S=+1/2}$, the level was populated by two
$\pi$-pulses transferring the population from
$\ket{S_{1/2},m_S=-1/2}$ via the state $\ket{D_{5/2},m_D=+1/2}$.
The transition frequencies were probed by Ramsey experiments with
the pulse lengths of the $\pi/2$ pulses adjusted to
$\tau=\unit[50]{\mu s}$. A Ramsey probe time of $T_R=\unit[1]{ms}$
yielded a minimum contrast of 86\% on the transition most
sensitive to magnetic field fluctuations. After the Ramsey
excitation, the state of the ion was detected by electron shelving
using a photo multiplier for fluorescence detection. A detection
time of \unit[2]{ms} led to a state discrimination efficiency well
above 99\%. This cycle was repeated a hundred times to obtain the
mean excitation probability. To infer the line center of a
transition, we switched the relative phase of the Ramsey pulses by
90\degree\ with a precision of better than $\pi \times 10^{-4}$
and repeated the experiment with reversed pulse order. This
arrangement is free of errors due to incorrectly set phases.

The dominant level shifts caused by electro-magnetic fields are
the linear Zeeman shift ($\propto$ \unit[10]{MHz}) induced
by the static magnetic field defining the quantization axis and
the electric quadrupole shift ($\propto$ \unit[10]{Hz}),
resulting from an interaction of the quadrupole moment of  the
$3d\ ^2D_{5/2}$ level \cite{roos} with static electric field
gradients caused by either the DC-trapping fields and possible spurious
field gradients of patch potentials. Both level shifts cancel out
when averaging over all frequency measurements of the six
transitions shown in Fig.\ \ref{term}(b).

\begin{figure}
\centering
\includegraphics[scale=0.45]{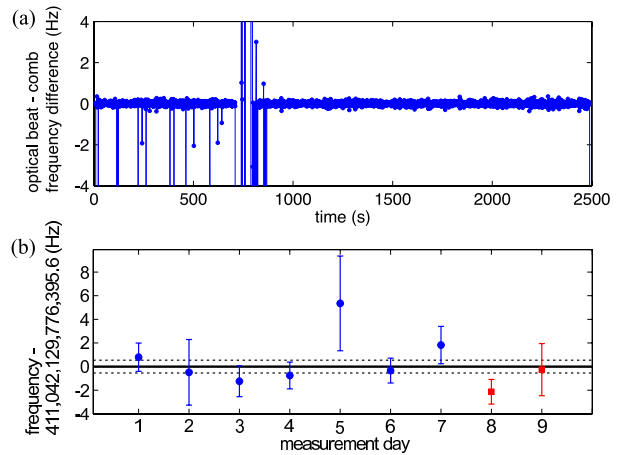}
 \caption{\label{tage} (a) Example for a data set where the direct optical beat signal of
both lasers was compared to the difference of the frequency comb
readings. This information was used to detect cycle slips of the
counters (gate time \unit[1]{s}) mainly caused by low signal-to-noise ratio. An error shows up as a
frequency difference larger than \unit[1]{Hz}. (b) Frequency
deviation during different days. The circles are mean values with
statistical error bars (1$\sigma$) for normal trap drive power
 of \unit[9]{W}. The solid line is the average of this data with
the statistical uncertainty indicated by dashed lines. The squares
show data obtained with the trap power reduced by a factor of
three.}
 \end{figure}

The six transitions were repeatedly probed in the order shown in
Fig.\ \ref{term}(b) in a measurement cycle taking \unit[22]{s},
and the mean excitation and the frequencies of all acousto-optical
modulators (AOM) for each setting were recorded. The measurement
results obtained from transitions 1 and 4
($\ket{m_S=-1/2}\leftrightarrow\ket{m_D=-1/2}$ and
$\ket{m_S=+1/2}\leftrightarrow\ket{m_D=+1/2}$) were used to infer
the current laser frequency relative to the ion and the magnetic
field strength. This information was fed back onto the frequency
of acousto-optic modulators (see Fig.~\ref{setup}) to compensate
for slow drifts of the reference cavity (AOM1) and the magnetic
field (AOM2). Together with a measurement of the quadrupole shift
as described in \cite{chwalla}, we were able to predict the
individual transition frequencies within the statistical
measurement accuracy of $\unit[\pm 0.4]{Hz}$ which is within a
factor of two of the expected quantum projection noise. The
individual transition frequency data combined with the Land\'e
g-factor of the $S_{1/2}$ level \cite{tommaseo} was used to
determine the g-factor of the $D_{5/2}$ level to be
g$_D$=1.2003340(3).

\begin{figure}
\includegraphics[scale=0.45]{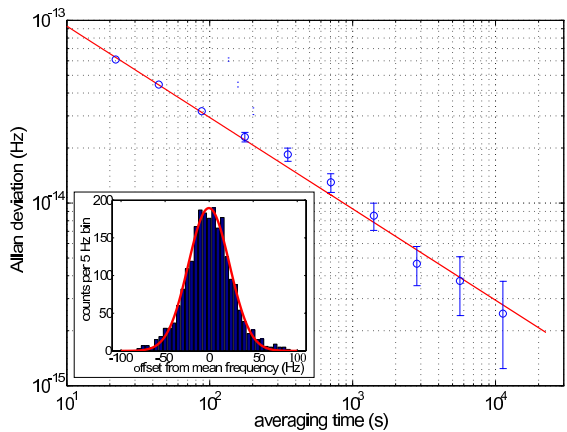}
\caption{\label{alan} Absolute frequency instability for the
comparison of $^{40}$Ca$^+$ against Cs. The data set consists of
2278 frequency measurement sets of all six transitions with an
effective gate time of 22 s. The inset shows a histogram of the
data which is consistent with a Gaussian fit of \unit[23(1)]{Hz}
standard deviation.}
 \end{figure}

A part of the probe laser light was sent to a frequency comb for a
measurement of its frequency. The frequency comb as well as all
radio frequency sources in the experiment, i.\ e.\ synthesizers
for AOMs tuning the laser into resonance with the ion or
cancelling fiber noise, were referenced to the transportable Cs
fountain atomic clock of LNE-SYRTE which has an accuracy of better
than 10$^{-15}$ \cite{bize} as was also confirmed by comparison of
the fountain clock with the ensemble of fountains in Paris before
and after the experiment. The second ion trap experiment (see
Fig.~\ref{setup}) served to check the validity of the frequency
comb readings. Three synchronized counters with a respective gate
time of \unit[1]{s} counted the beat note frequency $f_B$ of the
two lasers, the beat note frequency $f_1$ of the comb with the
first laser and the beat note frequency $f_2$ of the comb with the
second laser. If $||f_1-f_2|-f_B| \ge \unit[0.5]{Hz}$, the
measurement was removed (see Fig.\ \ref{tage}(a)). The recorded
measurement data was combined in the following way: the AOM
frequencies applied for each transition and the respective
frequency deviations inferred by the Ramsey experiments were
averaged together with the frequency comb data in the same time
window. The computers for data taking were synchronized to better
than \unit[0.1]{s}. The frequency of the $4s\ ^2S_{1/2}-3d\
^2D_{5/2}$ transition without correction for systematic shifts
could be determined to be 411\,042\,129\,776\,395.6(0.5)~Hz,
limited mainly by the stability of the frequency comb's repetition
rate. The average frequency per day is given in Fig.\
\ref{tage}(b). The Allan standard deviation for the data is shown
in Fig.\ \ref{alan}. The solid line is a fit with
$\sigma_y(\tau)=2.9(1)\times 10^{-13}\ \tau^{-1/2}$, suggesting
white frequency noise as the dominant noise source. The inset
shows a histogram of the deviation from the average frequency
which fits a Gaussian distribution with \unit[23(1)]{Hz} standard
deviation.

The influence of the most important effects on the transition
frequency with their magnitude and uncertainty is given in Tab.\
\ref{budget}. Since the linear Zeeman shift, the electric
quadrupole shift as well as the AC-Zeeman shift caused by
imbalanced currents in the trap electrodes cancel by the chosen
measurement scheme, the largest remaining shift is due to the
second order Zeeman effect which we determined to be
\unit[1.368(1)]{Hz} at a mean magnetic field of
\unit[3.087(2)]{G}. There was also a residual magnetic field drift
which on average was $2(1)\cdot 10^{-8}$~G/s. This lead to a
measurement offset of \unit[28(14)]{mHz} because this effect is
not completely canceled by averaging over the six transitions
without changing their order.

However, the largest uncertainty in the error budget stems from the AC-Stark shift induced by black body radiation. The ion is exposed to
thermal fields emanating from the surrounding vacuum vessel and
the ion trap which gets heated up by the applied RF power. These shifts can be calculated given the static scalar
polarizabilities of the $S_{1/2}$ and $D_{5/2}$ states
\cite{arora}. The actual trap temperature as a function of the
applied RF power can only be roughly estimated with
the help of a test setup similar to the ion trap we were using for
the frequency measurement. At the input power of \unit[9]{W}
that was applied for the frequency measurement, the trap heats up
to about \unit[150$\pm$50]{\celsius}. The trap only covers about
one third of the total solid angle while the rest is covered by
the surrounding vacuum vessel at a temperature of
\unit[25$\pm$2]{\celsius}, we estimate the black body shift to be
\unit[0.9(7)]{Hz}. The last two data points of Fig.\ \ref{tage}(b)
show measurements taken at a lower trap power (\unit[3]{W}) and
therefore lower trap temperature (\unit[60$\pm$20]{\celsius}). At
this level, no difference is apparent as the expected frequency
difference due to the change in black body radiation is on the
order of -0.4(4) Hz.

The Ramsey experiment leads to an AC-Stark shift
$\delta_{probe}=\delta_{AC}/(\frac{\pi T_R}{4\tau}+1)$, where
$T_R$ is the probe time and $\delta_{AC}$ the light shift during
the time $\tau$ of the Ramsey pulses which is mostly caused by
non-resonant excitation of dipole transitions \cite{hartmut}. For
our parameters, the expected shift is \unit[0.11(2)]{Hz}. The
light shift induced by the residual light of the resetting laser
at \unit[854]{nm} could be measured to be \unit[-0.005(4)]{Hz}.
The shifts by the cooling laser at \unit[397]{nm} and the
repumping laser at \unit[866]{nm} are estimated by a worst case
scenario: the residual amount of light which could be detected is
assumed to be focused tightly onto the ion. For the cooling laser
we assume a detuning of half a linewidth where the shift would be
maximum. The corresponding shifts and the uncertainties are well
below \unit[0.1]{Hz}. Excess micromotion induced by the trapping
field \cite{berkeland} is responsible for second order Doppler and
AC-Stark shifts. With the help of additional DC electrodes, we
compensate micromotion to modulation indices smaller than 1\%,
resulting in micromotion-related frequency errors of
\unit[0.1]{Hz} at most. Taking into account these systematic
shifts, the corrected value of the $4s\ ^2S_{1/2}-3d\ ^2D_{5/2}$
transition is $\nu_{Ca^+} = 411\,042\,129\,776\,393.2 (1.0)$~Hz
which corresponds to $2.4\cdot 10^{-15}$ relative uncertainty. To
our knowledge, this is the most accurate of any Ca$^+$ transition
frequency measurement so far \cite{privatecomm}.

It appears realistic that an experiment with $^{40}$Ca$^+$
especially designed for metrology could lead to an accuracy of better than
10$^{-16}-$10$^{-17}$. There are some advances necessary
to reach this level such as improved light attenuation when switching
off the lasers in order to eliminate AC Stark shifts. It is more
difficult to improve on the shift by black body radiation. For
this, a new trap would have to be designed with the possibility to
measure the trap temperature exactly. Additionally, operating the
trap in a cryogenic environment \cite{oskay} would dramatically
reduce the frequency uncertainty due to the black body shift.
Generally, a laser with enhanced stability could substantially
improve on the required measurement time for investigating
systematic effects by comparison of the two ion trap experiments.

\begin{table}
\caption{\label{budget} The systematic frequency shifts and their
associated errors in Hz and the fractional uncertainty in units of
$10^{-15}$.}
\begin{ruledtabular}
\begin{tabular}{l|c|c|c}
\bf{Effect} & \bf{Shift (Hz)} & \bf{Error (Hz)} & \bf{Fractional} \\
~ & ~ & ~ & \bf{err. }$(10^{-15})$\\ \hline
Statistical error & - & 0.5 & $1.2$\\ \hline
1$^{st}$-order Zeeman & 0 & 0.2 & $0.5$\\ \hline
Magnetic field drift  & 0.03 & 0.01 & $0.02$\ \\ \hline
2$^{nd}$-order Zeeman: & 1.368 & 0.002 & $0.005$\\
quantization field & ~ & ~  & ~\\ \hline
Electric quadrupole  & 0 & 0.2 & $0.5$\\ \hline
\underline{AC Stark shifts:}&~&~&~\\
Laser at \unit[729]{nm}& 0.11 & 0.04 & $0.1$\\ 
Lasers at 397, 866, & 0 & $<0.1$ & $<0.2$\\ 
and \unit[854]{nm} & ~&~ & $~$\\ 
Black body rad. & 0.9 & 0.7 & $1.7$\\ 
Micromotion  & 0 & 0.1 & $0.2$\\ \hline
2$^{nd}$-order Doppler & -0.001 & 0.001 & $0.002$\\ \hline
Cs uncertainty  & 0 & 0.4 & 1 \\ \hline
\bf{Total shift} & \bf{2.4} & \bf{1.0} & \bf{2.4}\\ 
\end{tabular}
\end{ruledtabular}
\end{table}

Our experimental setup is mainly dedicated to processing quantum
information, a field of research closely related to metrology: a
pair of entangled ions can be used for improving the
signal-to-noise ratio \cite{Bollinger96,Leibfried04} but also for
designing states which are immune to certain kinds of
environmental perturbations \cite{roos,chwalla}. States like
$\ket{S_{1/2},m_S}\ket{S_{1/2},-m_S}+\ket{D_{5/2},m_D}\ket{D_{5/2},-m_D}$
which are immune against magnetic field fluctuations to first
order could be used for generalized Ramsey experiments with probe
times limited only by the probe laser stability and spontaneous
decay of the metastable state. Preliminary tests with such states
showed significant phase errors caused by changes in the optical
path length in the electro-optical deflector used for steering the
beam to enable individual addressing of the ions. To overcome this
problem, beam steering of a strongly focused laser could be
avoided by generating high-fidelity entanglement \cite{jan} with a
laser beam collectively interacting with the ions, and combining
this wide beam with a second strongly focused laser beam inducing
phase shifts to make the ions distinguishable for carrying out
coherent operations on a single ion.

\begin{acknowledgments}
This work was supported by the Austrian "Fonds zur F\"orderung der
wissenschaftlichen Forschung", the Austrian Academy of Sciences,
IARPA, the European network SCALA, and the Institut f\"ur
Quanteninformation GmbH. The mobile fountain work is supported by
CNES, CNRS, LNE and r{\'e}gion Ile de France (IFRAF). The SYRTE is
a unit associated to CNRS UMR 8630. We thank P.~O.~Schmidt and
H.~H{\"a}ffner for fruitful discussions and a critical reading of
the manuscript.
\end{acknowledgments}


\begin{thebibliography}{99}
\bibitem{privatecomm}
Very recently, the transition frequency was also measured at NICT,
Tokyo, yielding $411\,042\,129\,776\,385\pm 18$~Hz, the value
being consistent with our result. Private communication by Ying
Li.
\end{thebibliography}
\end{document}